\documentclass[conference]{IEEEtran}
\IEEEoverridecommandlockouts
\pdfoutput=1
\usepackage{cite}
\usepackage{amsmath,amssymb,amsfonts}
\usepackage{algorithmic}
\usepackage{graphicx}
\usepackage{textcomp}
\usepackage{xcolor}
\usepackage{multirow}
\usepackage{tabularray}
\usepackage{todonotes}
\usepackage[frozencache=true,cachedir=minted-cache]{minted} 

\def\BibTeX{{\rm B\kern-.05em{\sc i\kern-.025em b}\kern-.08em
    T\kern-.1667em\lower.7ex\hbox{E}\kern-.125emX}}
\usepackage{enumitem}

\usepackage[colorlinks=true,linkcolor=blue,citecolor=blue,allcolors=blue]{hyperref}

\begin{document}

\title{CodeSift: An LLM-Based Reference-Less Framework for Automatic Code Validation}

\author{\IEEEauthorblockN{Pooja Aggarwal}
\IEEEauthorblockA{\textit{IBM Research}, India \\
aggarwal.pooja@in.ibm.com}
\\
\IEEEauthorblockN{Brent Paulovicks}
\IEEEauthorblockA{\textit{IBM Research}, US \\
ovicks@us.ibm.com}
\and
\IEEEauthorblockN{Oishik Chatterjee}
\IEEEauthorblockA{\textit{IBM Research}, India \\
oishik.chatterjee@ibm.com}
\\
\IEEEauthorblockN{Brad Blancett}
\IEEEauthorblockA{\textit{IBM}, US \\
blancett@us.ibm.com}
\and
\IEEEauthorblockN{Ting Dai}
\IEEEauthorblockA{\textit{IBM Research}, US \\
ting.dai@ibm.com} \\
\IEEEauthorblockN{Arthur De Magalhaes}
\IEEEauthorblockA{\textit{IBM}, US \\
arthurdm@ca.ibm.com}

\and
\IEEEauthorblockN{Prateeti Mohapatra}
\IEEEauthorblockA{\textit{IBM Research}, India \\
pramoh01@in.ibm.com}


}

\maketitle

\begin{abstract}

The advent of large language models (LLMs) has greatly facilitated code generation, but ensuring the functional correctness of generated code remains a challenge. Traditional validation methods are often time-consuming, error-prone, and impractical for large volumes of code. We introduce CodeSift, a novel framework that leverages LLMs as the first-line filter of code validation without the need for execution, reference code, or human feedback, thereby reducing the validation effort. We assess the effectiveness of our method across three diverse datasets encompassing two programming languages. Our results indicate that CodeSift outperforms state-of-the-art code evaluation methods.  Internal testing conducted with subject matter experts reveals that the output generated by CodeSift is in line with human preference, reinforcing its effectiveness as a dependable automated code validation tool.

\end{abstract}


\section{Introduction}

In today's software development landscape, the proliferation of large language models (LLMs) has vastly accelerated the pace and adoption of code generation. For instance, with the increase in IT deployments and cloud adoption, IT operations (ITOps) that are critical for maintaining reliable and resilient systems, can extend the use of AI and automation through code generation to reduce mean time to resolve incidents. When an incident occurs, Site Reliability Engineers (SREs) are tasked with diagnosing the fault and finding a resolution. Traditionally, SREs would typically write code scripts manually to carry out these resolutions. Recently, with the advent of LLMs, code generation capabilities are now used to assist in incident remediation and automation.

The widespread adoption of generated code brings a need for robust validation mechanisms to ensure its functional correctness before deployment in production systems. Deploying unvalidated code directly into these systems can lead to severe consequences, including performance degradation and critical system failures. While manual inspection is the gold standard solution, the sheer volume of generated code makes it labor-intensive and time-consuming.

Recent approaches to evaluate generated code either use reference codes to compare with or are based on runtime execution correctness~\cite{eghbali2023crystalbleu,ren2020codebleu, zhou2023codebertscore, dong2023codescore,du2023classeval,schafer2024testgen}. However, these approaches come with the need for availability of reference codes or test cases and are limited by scalability constraints with increased workload.


To address these issues, we propose a novel framework, called CodeSift, to act as a first-line filter for generated code evaluation. This framework is designed to validate the functional correctness of generated code using LLMs as evaluators. By harnessing the power of advanced natural language understanding, semantic comprehension, and code analysis capabilities offered by LLMs, 
 CodeSift offers an efficient solution to the challenge of code validation, significantly reducing the burden on human validators and streamlining the code validation process. Code snippets that pass the first-line validation can undergo further scrutiny by human experts or be tested in a controlled development environment, thus maximizing efficiency and minimizing the risk of deploying faulty code. 
 Our contributions are: 
 \begin{itemize}[leftmargin=*]
 \item We propose CodeSift, an automatic and effective validation framework for assessing the quality of generated code, eliminating the need for execution, reference codes, and test cases. 
 \item We contribute a novel Bash dataset consisting of $100$ unique tasks with corresponding test cases and example ground truth code. 
 \item We showcase CodeSift's efficacy as a primary filter for automated script validation across various datasets including opensource Python dataset such as HumanEval~\cite{chen2021codex} and MBPP~\cite{austin202mbpp} and our manually curated Bash dataset, through a comparative analysis with test case-based validation. 
 \item We present how CodeSift can serve as a valuable metric for ranking various code generation models, aiding in model selection and performance evaluation in the absence of ground truth reference codes or test cases. 
 \item Results from user studies indicate that CodeSift's evaluations closely align with human experts' judgments, demonstrating its reliability and effectiveness in assessing the functional correctness of generated code.
 \end{itemize}

\begin{figure*}[htbp]
  \centering
  \includegraphics[width=0.8\textwidth]{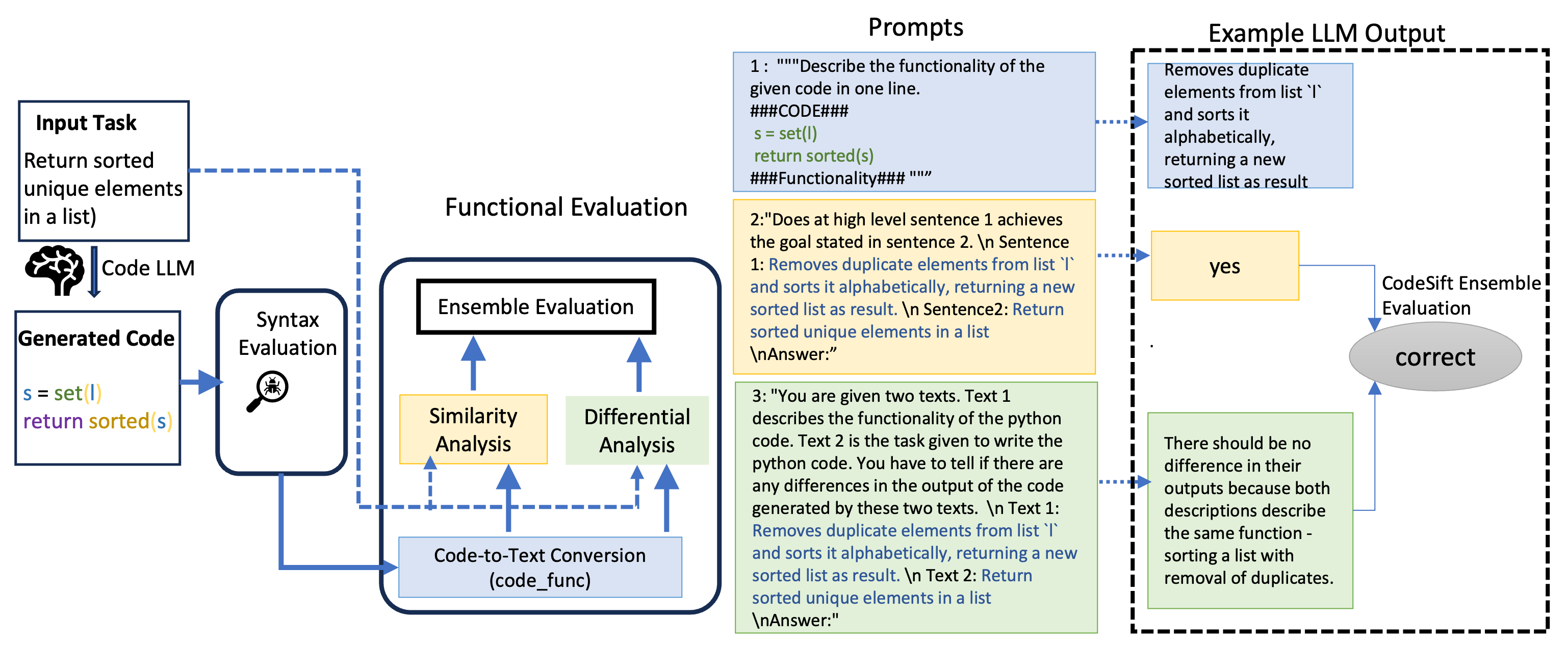} 
  \caption{CodeSift: Automated Script Validation System Diagram and Prompt Examples. Prompt 1 extracts code functionality, Prompt 2 assesses task-code similarity, and Prompt 3 analyzes differences."}
  \label{fig:sys_dia}
\end{figure*}

\section{Related work}
Many approaches for generated code evaluation have been proposed in the literature along four dimensions: (1) match-based (2) embedding-based, (3) execution-based, and (4) prompt-based.

Prior work such as BLEU, ROGUE, and ChrF assessed the quality of the generated code relying on token matching with the reference code.
BLEU score~\cite{papineni2022bleu} and ROUGE score~\cite{lin2004rouge} calculate the precision and recall of \textit{word} n-grams in the machine-generated code by comparing them to the reference code, respectively.
ChrF~\cite{popovic2015chrf} calculates the f-scores of the \textit{character} n-grams between the generated code and the reference code.
In programming languages, even unrelated pieces of code can share many common n-grams due to syntactic verbosity and coding conventions. Relying solely on n-gram matching fails to distinguish between similar code examples and those merely written using the same vocabulary.
CrystalBLEU~\cite{eghbali2023crystalbleu} addresses the issue of solely relying on n-gram matching by minimizing the noise caused by trivially shared n-grams, such as `(' and `,'. 
CodeBLEU~\cite{ren2020codebleu} assesses deep semantic similarities by incorporating weighted n-gram matching, syntactic AST matching, and semantic dataflow matching.

Embedding-based evaluation approaches for example, CodeBertScore~\cite{zhou2023codebertscore} measure the similarity between generated and reference code by summing the cosine similarities between their token embeddings and by incorporating contextual information.

Both match-based and embedding-based approaches demonstrate a poor correlation with human judgment or runtime execution validation. 
Moreover, these approaches only work when reference code is available, hindering their practicability.

Execution-based evaluation approaches evaluate code quality based on runtime execution correctness.
SPoC~\cite{kulal2019spoc} evaluates functional correctness using the pass@$k$ metric. For each problem, $k$ code samples are generated, and the problem is considered solved if any sample passes the unit tests.
Codex~\cite{chen2021codex} tackles the high variance issue in the pass@$k$ metric by generating $n \ge k$ samples per task. It counts the number of correct samples $c \leq n$ that pass unit tests and calculates the unbiased estimator.
APPS~\cite{hendrycks2021measuring} and CodeScore~\cite{dong2023codescore} evaluate functional correctness using the PassRatio metric, calculating the average fraction of test cases passed. 

Execution-based code evaluation approaches require running the generated code against a predefined set of test cases and comparing the output with expected results. While effective, they are inherently limited by scalability constraints. These limitations are mitigated by automated test case generation techniques~\cite{du2023classeval,schafer2024testgen,chen2022codet,codamosa2023,tufano2022test,mastropaolo2022using}. 
However, validating the generated test cases increases the workload, rendering the entire execution-based approach impractical.
Moreover, running model-generated code carries security risks and requires execution within a secure sandbox, which introduce additional technical complexity.



Prompt-based code evaluation approaches assessed the quality of generated code using LLMs with single answer grading,
pairwise comparison, reference-guided grading, and chain-of-thoughts~\cite{zheng2023judging,liu2023geval,zhuo2024icescore}.
They assign rating scores for evaluating the generated code by comparing it with the given task description or/and with the reference ground truth code.
They conduct text-to-code or code-to-code comparisons.
Inspired by those methods, in contrast to the above approaches, we use LLMs as a first-line filter to validate the functional correctness of the generated code by translating code into text dimension first and conducting the text-to-text comparison.


\section{Bash Dataset}
\label{sec:bash_dataset}

One of the emerging usages of CodeLLMs in AIOps is to automate incident remediation using automatic script generation for recommended actions. Bash is widely used by SREs for incident remediation scripting, primarily due to the prevalence of Linux-based systems. Hence it is crucial to evaluate the performance of CodeLLMs on the task of generating bash scripts. To evaluate LLMs on the task of code generation, people calculate execution accuracy (pass@$k$)~\cite{evaluatepassk2021} for benchmark datasets across different languages such as Python, Java, and Go~\cite{humanevalx}. These datasets contain the problem statement and associated test cases. When the given code passes all the test cases for the problem, it is marked as correct. Creating test cases for Bash scripts is challenging because Bash commands frequently alter the system. Therefore, verifying the code's success in completing the given task requires careful inspection of the system's state.

We create a new benchmark dataset consisting of 100 tasks for evaluating LLMs on generating bash scripts. The evaluation of bash scripts typically consists of the following steps: 
1) Prologue: This step consists of creating a container along with the necessary prerequisites required to evaluate the code. For example, if the task is to copy a file from directory1 to directory2, a container is created containing  directory1 with the file and directory2. 
2) Code Execution. 
3) Epilogue: In this step, we check for any unnecessary system changes 
4) Evaluation: We first check if the code was executed without any errors. Then we check if the code was able to fulfill the intended task. 
5) Cleanup: The container is closed and the execution result is returned.

\section{CodeSift Framework}


In this section, we outline the methodology of our code validation framework. Our objective is to check whether the generated code is correct and its functionally aligned with the desired behavior specified by the given task\footnote{In this context, "task" refers to programming problems.}. To accomplish this, CodeSift acts as a first-line filter to identify functionally incorrect code. This approach enables automated validation of code functionality without the need for execution or reference code, thereby reducing the validation effort and accelerating the adoption of generated code in real-world applications. The process is illustrated in Figure~\ref{fig:sys_dia}. 
CodeSift consists of two primary evaluation components: (1) \textit{syntax correctness evaluation} and  (2) \textit{semantic correctness evaluation}. 

In the Syntax Correctness phase, we utilize pre-built syntax checkers such as ShellCheck~\cite{shellcheck} and PyLint~\cite{pylint} to detect any syntactic errors in the generated code. If an error is detected, the LLM model is prompted with both the error message and the previously generated code. Our observations indicate that while the majority of generated code is syntactically correct, in cases where errors arise, LLM models can often rectify them when provided with the error message.

After successfully passing syntax evaluation, the code proceeds to undergo semantic correctness evaluation, which comprises of three main phases. In each of these phases, we utilize the same LLM. 
 Below, we elaborate each phase of semantic validation.
\begin{itemize}[leftmargin=*]
    \item \textbf{Code-to-functionality}: 
   In this phase, the generated code is transformed into a text representation, referred to as \textit{code-func}, which encapsulates the core functionality of the code. To accomplish this, we leverage a pre-trained LLM, capable of generating the primary purpose and operational logic corresponding to the provided code. Prompt 1 in Figure~\ref{fig:sys_dia} is used for this phase.
\item   \textbf{Similarity Analysis}: 
    Next, we utilize the same pre-trained LLM to assess the semantic similarity between the \textit{code-func} and the task description. The LLM is prompted to determine whether the \textit{code-func} can accomplish the intended behavior specified in the task. Prompt 2 in Figure~\ref{fig:sys_dia} is utilized to assess the similarity between the task and \textit{code-func}. If the similarity analysis indicates that both the \textit{code-func} and the task achieve the same goal, then the code will be labeled as functionally correct in this phase.
    \item  \textbf{Difference Analysis}:
In the third phase, we conduct a difference analysis by instructing the same LLM model to identify and examine any discrepancies between the \textit{code-func} and the task description. Prompt 3, shown in Figure~\ref{fig:sys_dia}, is utilized to discern differences within both texts. This helps in identifying semantic variations between the two texts, enabling the detection of potential discrepancies or inconsistencies between the expected and generated output. 
If the difference analysis indicates that the \textit{code-func} and task produce identical outputs and exhibit no discrepancies, the code will be labeled as functionally correct during this phase. This phase complements the similarity analysis phase by providing additional insights into the functional correctness of the code, ultimately contributing to a more comprehensive evaluation.
\item \textbf{Ensemble Synthesis}:
The ensemble approach involves integrating the outputs of both similarity analysis and difference analysis to arrive at a more comprehensive evaluation of the generated code. The generated code is marked as correct by the CodeSift method only if the similarity analysis indicates that the \textit{code-func} and task are similar, and the difference analysis finds no disparities between them. Consequently, we synthesize the results from both analyses to comprehensively evaluate the generated code, thereby determining its functional correctness. If either analysis indicates a deviation from the task, the code is labeled as functionally incorrect. It's important to note that although the model receives identical inputs for both prompts, its focus varies depending on the objective. When assessing alignment with the task, the model emphasizes more on similarities and sometimes might disregard subtle differences. Conversely, when tasked with explicitly identifying differences, the model effectively does so. Therefore, capturing similarities and differences explicitly is crucial for accurately labeling the generated code, achieved by combining the outputs of both analyses.
\end{itemize}

\section{Evaluation}
Here, we present the data for evaluation and present the performance of our evaluation framework using various models and baselines.

\begin{table*}[]
\caption{\footnotesize {Accuracy of various LLM-based Code Evaluation methods. The best performance is \textbf{bold}. }}
\label{tab:main_results}
\centering
\scriptsize
\begin{tblr}[]{
        hline{1,4,Z}={1pt},
        vline{1,4,Z}={1pt},
        vline{8,9}={solid},
        hline{2,6,9,12}={solid},
        rows={m,rowsep=0pt},
        colspec={X[-1,l]X[-1,l]X[-1,l]X[-1,c]X[-1,c]X[-1,c]X[-1,c]X[-1,c]X[-1,c]X[-1,c]X[-1,c]X[-1,c]},
        cell{1}{1} ={r=3,c=3}{c},
        cell{1}{4}={c=9}{c},
        row{2} = {font=\bfseries},
        cell{2}{4,9}={c=4}{c},
        cell{4}{1}={r=2}{m},
        cell{6}{1}={r=3,c=2}{m},
        cell{even[4-16]}{3-Z}  = {bg=gray!10},
}
\textbf{Methods with Code Evaluation Models} &  &  &\textbf{Dataset with Code Generation Models} &&&& &&&\\
&&& HumanEval (Python) & & & & Bash & MBPP+ (Python) &\\
 &&                                                      & Starcoder & GPT3.5 & Codellama & Mistral & Codellama  & Starcoder  & GPT3.5 & Codellama & Mistral\\
\textbf{Baseline} 
                     & \textbf{Reference Grading~\cite{zheng2023judging}} & Mistral   & 70.79     & 34.70  & 57.50     & 59.82   & 64.2       & 50.63      & 29.57  & 40.35     & 55.14  \\
                       & \textbf{ICE-Score~\cite{zhuo2024icescore}} & Mistral   & 64.6        & 50.4     & 58.8        & 51.1      & 61.9       & 64.1         & 51.6     & 50.8      & 55.1     \\
\textbf{CodeSift }& & Mistral     & \textbf{72.92}     & \textbf{65.3}   & \textbf{67.62}     & \textbf{66.76}   & \textbf{67.1}       & 62.4       & 56.6   & \textbf{62.4}      & 57.3   \\
                                        & & Mixtral     & 67.56     &  58      & 65.42     & 60.30   & 61.6       &     \textbf{66.9}       &  \textbf{71.0}      & 51.2      &    \textbf{59.6}    \\
                                   & & Llama2-Chat     & 62.85     & 51.7   & 54.63     & 55.60   & 59.7       & 60.4       & 66.4   & 59.6      &  58.8  \\
\end{tblr}
\end{table*}

\begin{table*}[]
\caption{\footnotesize {Accuracy of various components of CodeSift framework.}} \label{tab:comp}
\label{tab:main_results}
\centering
\scriptsize
\begin{tblr}[]{
        hline{1,4,Z}={1pt},
        vline{1,4,Z}={1pt},
        vline{8,9}={solid},
        hline{2,7,10,12}={solid},
        rows={m,rowsep=0pt},
        colspec={X[-1,l]X[-1,l]X[-1,l]X[-1,c]X[-1,c]X[-1,c]X[-1,c]X[-1,c]X[-1,c]X[-1,c]X[-1,c]X[-1,c]},
        cell{1}{1} ={r=3,c=3}{c},
        cell{1}{4}={c=9}{c},
        row{2} = {font=\bfseries},
        cell{2}{4,9}={c=4}{c},
        cell{4}{1}={r=3}{m},
        cell{7}{1}={r=3,c=2}{m},
        cell{even[4-16]}{3-Z}  = {bg=gray!10},
}
\textbf{Methods with Code Evaluation Models} &  &  &\textbf{Dataset with Code Generation Models} &&&& &&&\\
&&& HumanEval (Python) & & & & Bash & MBPP+ (Python) &\\
 &&                                                      & Starcoder & GPT3.5 & Codellama & Mistral & Codellama  & Starcoder  & GPT3.5 & Codellama & Mistral\\

\textbf{CodeSift's Similarity Analysis}& & Mistral      & \textbf{71.40}     & 67.2   & \textbf{73.29}     & 56.34   & \textbf{67.6}       & 63.9       & 70.6   & 64.9      & 54.3   \\
                                       & & Mixtral      & 62.92     & 68       & 64.20     & 54.63   & 66.4       &   \textbf{65.9}         &   74.0     & \textbf{67.4}      &   \textbf{59.3}     \\
                                   & & Llama2-Chat      & 52.15     & \textbf{71.1}   & 54.57     & 41.03   & 63.1       & 63.9       & 76.9   & 64.1      & 51.3   \\
\textbf{CodeSift's Difference Analysis}& & Mistral      & 61.03     & 65.1   & 55.48     & \textbf{60.1}   & 67.4       & 60.9       & 56.3   & 63.1      & 56.8   \\
                                       & & Mixtral      & 60.36     &  59      & 61.89     & 56.15   & \textbf{59.7}       &   64.1         &   \textbf{77.0}     & 51.3      &     54.3   \\
                                    & &Llama2-Chat     & 57.68     & 51.8   & 52.68     & 54.14   & 55.9       & 58.1       & 68.4   & 61.1      & 58.6   \\
\end{tblr}
\end{table*}

\begin{table*}[]
\caption{\footnotesize {Precision of various LLM-based Code Evaluation methods.}}
\label{tab:precision_results}
\centering
\scriptsize
\begin{tblr}[]{
        hline{1,4,Z}={1pt},
        vline{1,4,Z}={1pt},
        vline{8,9}={solid},
        hline{2,6,9,12}={solid},
        rows={m,rowsep=0pt},
        colspec={X[-1,l]X[-1,l]X[-1,l]X[-1,c]X[-1,c]X[-1,c]X[-1,c]X[-1,c]X[-1,c]X[-1,c]X[-1,c]X[-1,c]},
        cell{1}{1} ={r=3,c=3}{c},
        cell{1}{4}={c=9}{c},
        row{2} = {font=\bfseries},
        cell{2}{4,9}={c=4}{c},
        cell{4}{1}={r=2}{m},
        cell{6}{1}={r=3,c=2}{m},
        cell{even[4-16]}{3-Z}  = {bg=gray!10},
}
\textbf{Methods with Code Evaluation Models} &  &  &\textbf{Dataset with Code Generation Models} &&&& &&&\\
&&& HumanEval (Python) & & & & Bash & MBPP+ (Python) &\\
&&  & Starcoder & GPT3.5 & Codellama & Mistral & Codellama  & Starcoder  & GPT3.5 & Codellama & Mistral\\
\textbf{Baseline} 
& \textbf{Reference Grading~\cite{zheng2023judging}}             & Mistral   &   62.4      &   91.6   &   76.6      & 34.5    & 92.5          &   76.1      &   91.1   &   84.8      & 77.1 \\
& \textbf{ICE-Score~\cite{zhuo2024icescore}}                     & Mistral   &   45.5      &   71.1   &   56.4      & 38      & 59.8          &   67.3      &   80.4   &   66.1      & 54.8 \\
\textbf{CodeSift }& & Mistral   &   57.1      &   75.9   &   72.9      & 54.5    & 73.1          &   63.9      &   83.4   &   69.3      & 55.3\\
                                        & & Mixtral     &  48.7 & 77.0 & 75.3 & 57 & 75.3 & 66.1 & 85.0 & 74.2 & 56.4\\
                                      & & Llama2-Chat   &  41 & 73.1 & 52.8 & 73.1 & 69.2 & 66.5 & 83.2 & 41.1 & 55.9 \\
\end{tblr}
\end{table*}

\begin{table*}[ht]
\caption{Examples illustrating instances of successful performance and instances of failure of the CodeSift method.}
\label{tab:examples}
\scriptsize
\centering
\begin{tblr}{
        hline{1,2,Z}={1pt},
        hline{2-9}={solid},
        colspec={X[2,l]|X[2,l]|X[2,l]|X[1.5,l]},
        rows={m,rowsep=0pt},
        row{1}= {c},
    }
\textbf{Input Task}       & \textbf{Generated Code}   & \textbf{Functionality (code-func)} & \textbf{CodeSift's Evaluation} \\
You are given a list of integers. Write a function next\_smallest() that returns the 2nd smallest element of the list. Return None if there is no such element.
&
\begin{minipage}{0.2\columnwidth}
\begin{minted}{python}
if len(lst) < 2:
    return None     
else:         
    return sorted(set(lst))[1]
\end{minted}
\end{minipage}
& The code checks whether a list `lst' has less than two elements, and returns `None' if so; otherwise, it sorts a set made from the list's elements, then returns the second element of the resulting sorted list.
& {The python code \textbf{does not provide any information about how to find the second smallest element.}}  \\
Please write a function to efficiently compute the n-th element of the fib4 number sequence.  \textbf{Do not use recursion}.    
&
\begin{minipage}{0.5\columnwidth}
\begin{minted}{python}
if n == 0:
    return 0
...
if n > 3:
    return fib4(n-1) + fib4(n-2) + fib4(n-3) + fib4(n-4)
\end{minted}
\end{minipage}
& This \textbf{recursive} function calculates ... until reaching base cases and sums up their returned results. 
& The \textbf{code being described is a recursive} function for fib4 whereas in \text{task it is described is a non-recursive}  \\ 
Given length of a side and high return area for a \textbf{triangle}. 
&
\begin{minipage}{0.3\columnwidth}
\begin{minted}{python}
return a * h / 2
\end{minted}
\end{minipage}
& This function computes and returns the area of a \textbf{rectangle} with base `a' and height `h'. 
& 
The generated \textit{code-func}  is incorrect as it asserts that the formula is for a rectangle instead of a triangle. \\ 
Given a string representing a space separated lowercase letters, return a dictionary
    of the letter with the most repetition and containing the corresponding count.
&
\begin{minipage}{0.3\columnwidth}
\begin{minted}{python}
if test == '':
        return {}
else:
    ...
            count[i] = 1
    return count
\end{minted}
\end{minipage}
& This function takes a string as an argument, ... iterates through \textbf{each word} ...its value gets incremented by 1 ... Finally, this dictionary is returned.
& 
The code generates a frequency distribution dictionary \textbf{for every unique words} present in the input string, while Tasks aims to find \textbf{only the word(s) having maximum repetitions} \\ 

Print total space used on device overlay
&
\begin{minipage}{0.3\columnwidth}
\begin{minted}{bash}
#!/bin/bash
df -h | grep overlay 
| awk '{print \$5}'
\end{minted}
\end{minipage}
& This bash script uses the 'df' command ... and finally, pipes the filtered output to the 'awk' command which extracts and prints the 5th column (representing the \textbf{percentage of used space}).
& 
The given bash script \textbf{does not print the total space used} on the device overlay. It only prints the \textbf{percentage of used space} on the overlay filesystems.\\ 

Copy all .txt files from directory dir1 to dir2 in the current directory
&
\begin{minipage}{0.3\columnwidth}
\begin{minted}{bash}
#!/bin/bash
cp -r /dir1/*.txt /dir2
\end{minted}
\end{minipage}
& This Bash script uses the 'cp' command to recursively copy all '.txt' files from directory '/dir1' to directory '/dir2'.
& 
 Yes, the given code description achieves the goal stated in the task.\\ 

Count the number of lines in a.txt file in current directory
&
\begin{minipage}{0.3\columnwidth}
\begin{minted}{bash}
#!/bin/bash
wc -l < a.txt
\end{minted}
\end{minipage}
& This Bash script uses the 'wc' (word count) command with its '-l' option to determine and print the number of lines in the file named "a.txt".
& 
Yes, this bash script achieves the goal stated in the task by counting the number of lines in the specified file using the wc command with the -l flag.\\ 

\end{tblr}
\end{table*}

\subsection{Experimental Setup}
\label{sec:exp_setup}
\subsubsection{Dataset}
We evaluate our framework on three datasets, including  \textbf{HumanEval}~\cite{chen2021codex},  \textbf{MBPP+}~\cite{mpbbplus}, 
and \textbf{Bash} (see Section \ref{sec:bash_dataset}).  The HumanEval dataset consists of $164$ Python programming problems while MBPP+, which is a refined subset of the original MBPP dataset \cite{austin202mbpp} by Evalplus \cite{evalplus}, contains $399$ python problems.

We analyze the performance of our framework on the code generated by four prominent models: Starcoder \cite{li2023starcoder}, Codellama\_34B \cite{roziere2023code}, ChatGPT \cite{OpenAI2023ChatGPT}, and Mistral\_7B \cite{jiang2023mistral}. 
For the HumanEval dataset, we increase the size for evaluation by sampling $10$ solutions for each problem from these models, using a temperature of $0.2$ (except for ChatGPT for which the temperature was set to $0.8$). This results in a total of $1640$ task-code pairs for each model. Similarly, for Bash, we generate $10$ scripts for 100 tasks to get a total of 1000. For the MBPP+ dataset, we use greedy decoding to generate $399$ task-code pairs.  
\subsubsection{Task}
We evaluate CodeSift on the following two tasks: i) determining the functional correctness of the given code, ii) its utility as a filter in code generation pipeline to reduce incorrect code shown to the user.  For the first task, we use accuracy as the metric to compare the performance of CodeSift with baseline methods. Establishing ground truth by marking task-code pairs as correct only if all associated test cases pass. We then calculate the accuracy by comparing the output of CodeSift with this ground truth. For the second task, we report the \% of times correct code was shown to the user post-filtering by CodeSift and baseline methods. 

\subsubsection{Models}
We evaluate the performance of three models: Llama2-Chat\_70B \cite{touvron2023llama}, Mistral\_7B \cite{jiang2023mistral}, and Mixtal\_8x7B~\cite{jiang2024mixtral}, in CodeSift's framework for functionality generation, similarity analysis and difference analysis.
All the models use the sampling decoding method with $0.6$ temperature and $1.2$ repetition penalty. 

\subsubsection{Baselines}
We compare CodeSift with two baseline methods: \textbf{ICE-Score} and \text{Reference Grading}. ICE-Score determines code correctness by directly comparing it with the specified task using a Large Language Model~\cite{zhuo2024icescore}. Reference Grading (as described in the LLM-as-a-judge framework~\cite{zheng2023judging}) also considers a correct reference code in addition to the code and task to improve accuracy. For both methods, we use prompts specified in their respective frameworks. ICE-Score assigns scores from 0 to 4, where 4 indicates functional correctness, while Reference Grading assigns a score of 10 for perfect alignment with the task. We classify codes as correct based on these scoring criteria.

\subsection{Results}
\label{sec:results}

The results of our experiments are structured around several key research questions (RQ). 

\textit{RQ1: How effective is CodeSift at assessing the accuracy of generated code?}

CodeSift's effectiveness in assessing the accuracy of generated code is evident from the comparison presented in Table~\ref{tab:main_results}.  CodeSift provides insights into the fact that LLMs excel in text-to-text comparisons rather than text-to-code evaluations. 
When utilizing Mistral as the backbone, CodeSift outperforms both baselines for the HumanEval and Bash datasets, while achieving superior performance with Mixtral for the MBPP dataset. CodeSift demonstrates superior performance over ICE-Score across all datasets, highlighting its effectiveness in text-to-text comparisons. Notably, in the Bash dataset, ICE-Score tends to label most codes as correct. However, even with strict scoring criteria, which categorize codes as incorrect with partial inaccuracies, ICE-Score overlooks certain differences. In contrast, by converting the code to text that captures its core functionality, the same LLM model can discern these differences. Consequently, while ICE-Score exhibits high recall, its precision is $13\%$ lower compared to CodeSift. Moreover, in the Python datasets (HumanEval and MBPP), CodeSift outperforms ICE-Score by an average of $8\%$. This observation underscores the significance of the LLM's prompt formulation, emphasizing the effectiveness of text-to-text comparison over text-to-code comparison.
Interestingly, despite the availability of reference code, the reference-based grading approach does not outperform CodeSift, highlighting the challenges of code-to-code comparison encountered by large language models (LLMs). This observation contradicts the intuitive assumption that having access to more information, such as reference codes, would be advantageous. However, in the realm of code, where multiple methods can achieve the same functionality, this notion does not hold. We notice that Reference Grading tends to perform poorly in cases where the generated code is produced via GPT3.5.  The ratings assigned by Reference Grading for these scenarios typically hover around $7$, indicating partial correctness. Hence it is labeled as incorrect. Consequently, even though the generated code may be functionally accurate, it is labeled as incorrect by Reference Grading due to its scoring criteria. Note that this issue with Reference Grading does not exist with other datasets since its performance remains comparable to CodeSift, suggesting that the overall scoring criteria of Reference Grading is appropriate.

\textit{RQ2: Is there alignment between the assessments made by CodeSift and human preference in evaluating the correctness of generated code?}
To evaluate the practical utility of CodeSift, it was deployed in an internal offline code generation pipeline for catalog creation of automation scripts for a widely-used AI observability platform. We conducted a user study involving 3 Subject Matter Experts (SMEs) who were tasked with generating and evaluating bash scripts using CodeSift. The experts were asked to assess the code functionality generated by the CodeSift along with the validation output. Feedback was received on $105$ instances. The SMEs agreed with the code functionality output $78\%$ of the time and agreed with the validation output $83\%$ of the time. This study shows the effectiveness of CodeSift as a code evaluator in real-world applications but also helps us evaluate the quality of the different key components of CodeSift - functionality generation and functional validation, through user feedback.

\textit{RQ3: How does CodeSift's performance change with different LLMs as evaluators, and what factors influence performance variations among LLMs?} 

Our experiment results reveal performance disparities among three LLMs as evaluators: Mistral, Mixtral, and Llama2-Chat, as demonstrated in Table~\ref{tab:main_results}. Specifically, the Mistral model consistently outperforms the other models across the HumanEval and Bash datasets. However, in the MBPP dataset, Mixtral's performance is superior. Notably, when considering precision (in Table~\ref{tab:precision_results}), CodeSift with Mixtral ensures that users are presented with fewer incorrect cases, thereby enhancing the overall user experience. While reference grading exhibits high precision, it suffers from low recall, resulting in few cases being labeled as correct. In contrast, CodeSift (with Mixtral), compared to ICE-Score, which also does not consider reference code, demonstrates $5\%$ to $12\%$ better precision across most scenarios, except in the case of Starcoder-MBPP.
Additionally, when considering the performance of the individual phases of CodeSift, namely similarity analysis and difference analysis, we observe similar performance between Mixtral and Mistral as shown in Table~\ref{tab:comp}. However, Mistral's similarity analysis can sometimes effectively highlight dissimilarities between the code functionality and task compared to other models. For instance, in the first example in Table~\ref{tab:examples}, only similarity analysis using Mistral model accurately identified dissimilarity between the task and \textit{code-func} whereas using other models for similarity analysis labeled it similar. 
We observed that the evaluator LLMs exhibit no bias, as evidenced by Mistral LLM outperforming others on codes generated by the Mistral model. This indicates the model's ability to detect errors in generated code, even when the code generation and evaluation models are the same. 
 The accuracy of CodeSift with Llama2-Chat as the backbone model is generally lower than CodeSift with Mistral and Mixtral. However, in cases where code is generated by the GPT3.5 model, CodeSift-Llama's similarity analysis phase performs notably well due to the higher execution accuracy of the GPT3.5 model. Consequently, CodeSift-Llama tends to label most generated code as correct, potentially overlooking discrepancies between the code functionality and the task requirements. This may result in lower overall accuracy, particularly in detecting incorrect code instances.

\textit{RQ4: Can CodeSift effectively detect functional errors in the generated code, including logical errors, syntax errors, and missing functions?} 

One interesting finding of our analysis is that LLMs can sometimes detect issues with code validity that current validation schemes (such as unit tests) may fail to capture.
This can be seen in the second example in Table~\ref{tab:examples}. Here the task mentions ``\textit{do not use recursion}''. The test cases cannot capture this aspect and hence the generated code using recursion is deemed correct by the execution evaluation. However, CodeSift accurately detects the use of recursion in the generated code, correctly determining that it does not fulfill the intended task by
stating that ``\textit{No ... To efficiently compute the n-th element of the fib4 sequence without using recursion, one would need to implement an iterative solution instead.}"
and suggests using ``\textit{iterative solution instead}''. For bash, in the fifth example of Table~\ref{tab:examples}, CodeSift was able to recognise that the $5^{th}$ row of df command's out represents used $\%$ rather than the actual value and was able to correctly mark it as incorrect. 
One of the reasons for CodeSift's failure to evaluate the code correctness is due to incorrect functionality generation. 
For example, in the third entry of Table~\ref{tab:examples}, CodeSift fails to provide the correct output due to a discrepancy between the \textit{code-func} and the generated code. This leads to false negatives and hinders CodeSift's overall accuracy. Another source of inaccuracy occurs when the generated code is very close to the actual intended task with minor discrepencies. For example, in the sixth entry of table~\ref{tab:examples}, the LLM is unable to determine that the task has asked to copy from $dir1$ to $dir2$ ( both directories are in the same current directory) where as the code copies from $dir1$ to $dir2$ which results in it claiming that the code is correct whereas the execution fails due to incorrect directory.

\section{Conclusion and Future Work}

We introduced a novel approach using LLMs for automatic code evaluation, demonstrating its usefulness in the absence of reference code and test case. Our experimental results demonstrate the effectiveness of CodeSift across various datasets and programming languages, outperforming baseline approaches such as ICE-Score and reference-based grading. Notably, CodeSift's ensemble approach, incorporating both similarity and difference analysis phases, yields the most reliable outcomes, providing a comprehensive assessment of code functionality.  In the future, we plan to refine and expand the capabilities of CodeSift by exploring its performance across a broader range of programming languages. Additionally, we aim to enhance the interpretability and transparency of the framework's decision-making process. Explaining the rationale behind the framework's assessments in a more intuitive and human-understandable manner can increase trust in its recommendations, facilitating its adoption in real-world software development workflows. We'll also leverage explanations generated by the framework to offer feedback to the code generation model, ensuring better alignment with task specifications.
\section{Limitations}

Our framework heavily relies on the functionality generated for a given code snippet. If this phase fails to capture the essential elements of the code, there is a higher likelihood of incorrectly labelling the generated code. Although Mistral and Mixtral models demonstrate proficiency in capturing essential functionality, instances of incorrect code functionality generation can still occur. Also since the outputs of the models are verbose, it is not always possible to automatically detect the correctness of the code from the similarity/difference analysis.  Additionally, the explanations provided by the similarity and difference analysis phases may not always be entirely accurate upon manual inspection. Future improvements to these explanations could enhance the framework's efficacy and provide valuable feedback to the code generation model. 






\bibliographystyle{IEEEtran}
\bibliography{conference_101719}

\end{document}